\documentclass[11pt]{article}
\usepackage{amssymb,amsfonts,amsgen,amsmath,latexsym}

\def\i{\begin{eqnarray}}\def\f{\end{eqnarray}}
\def\non{\nonumber}\def\del#1#2{\frac{\partial{#1}}{\partial {#2}}}
\def\vep{\varepsilon}\def\q{\quad}\def\C{{\mathbb C}}\def\H{{\cal H}}
\def\a{\alpha}\def\b{\beta}\def\R{{\mathbb R}}
\def\vd{\delta}\def\th{\theta}\def\tphi{\tilde{\phi}}\def\P{{\mathbb P}}
\def\lra{\longrightarrow}\def\tL{\tilde{L}}\def\tG{\tilde{G}}

\begin{document}

\begin{center}
{\Large\bf
On the Differential equations of the characters \\
\bigskip
for the Renormalization group} \\ 
\bigskip\bigskip
Masato Sakakibara \\
\smallskip
\textit{\small Department of Physics, University of Tokyo, \\ 
\smallskip
Tokyo 113-0033, Japan \\
\smallskip
sakakiba@monet.phys.s.u-tokyo.ac.jp}
\end{center}

\begin{abstract}
Owing to the analogy between the Connes-Kreimer theory of the renormalization
and the integrable systems, 
we derive the differential equations of the unit mass for the 
renormalized characters $\phi_+$ and the counter term $\phi_-$.
We give another proof of the scattering type formula of $\phi_-$. 
The differential equation of $\phi_-$ of the coordinate $\vep$ on $\P^1$ is also given. 
The hierarchy of the renormalization groups is defined as the integrable systems.
\end{abstract}

\section{The Conne-Kreimer theory}

We start on the same setting\footnote{Our notations
$\phi,\phi_\pm,\mu$ in this note
correspond to $\gamma,\gamma_\pm,\mu^{\frac{1}{2}}$ respectively in Ref. \cite{ck1,ck2}.}
 with the Connes-Kreimer's papers \cite{ck1,ck2}.
Let $\H$ be a Hopf algebra of the 1PIs of $g\phi^3$-theory.
Let $G$ be a Lie group of the characters of $\H$ and $L$ the Lie algebra
of derivations.
The product of $\phi_1,\phi_2\in G$ is given by
\i\label{00} 
(\phi_1\phi_2)(X)=\langle\phi_1\otimes\phi_2,\Delta(X)\rangle\q{\rm
for\; any}\, X\in\H. \f
The inverse and a unit are defined by $\phi^{-1}(X)=\phi(S(X))$ 
and $1(X)=\vd_{X1}$. The Lie bracket of $\vd_1,\vd_2\in L$ is defined as 
$([\vd_1,\vd_2])(X)=\langle \vd_1\otimes\vd_2-\vd_2\otimes\vd_1,X\rangle$ for any $X\in\H$.
The dual space $\H^\ast$ is an algebra with the
product (\ref{00}).
We add the element $Z_0$ to $L$ such that $\th_t:={\rm Ad}\,e^{tZ_0}$ is 
the grading\footnote{We should not
confuse the degree $\phi$ as the loop group and the algebra $\H^\ast$.}
of $G$. We define $\tG:=G\rtimes_\th\R$ and $\tilde{L}:=L\oplus\C Z_0$.

We consider the loop groups\footnote{In this note, we use same notation $\tG$ (resp. $\tL$) with
its loop group (resp. algebra).}
of $\tG$ and the loop algebra of $\tL$. Let $\vep$ be the
coordinate of $\P^1$ for the loop group which corresponds to the parameter of the 
dimensional regularization.
The Birkhoff decomposition in the sense of Ref. \cite{ck1,ck2} 
divides $\tG$ into $\tG=\tG_-\tG_+$, i.e., any character $\phi\!\in\!\tG$
decomposes uniquely as $\phi=\phi_-^{-1}\cdot\phi_+$ with the condition 
$\phi_-=1$ at $\vep=\infty$. 
The Lie algebra $\tL$ decomposes into $\tL=\tL_-\oplus\tL_+$. 
Keeping in mind the Feynman rules, we assume that $\phi,\phi_\pm$ depend on
the coupling constant $g=:e^{x}$, the unite mass $\mu=:e^t$ and
$\vep\in\P^1$ such that 
\i \phi=\phi(x+\vep t,\vep),\q \phi_-=\phi_-(x,\vep),\q
\phi_+=\phi_+(x,t,\vep). \f

\section{The differential equations of the unit mass}

The adjoint action of $e^{t\vep Z_0}$ on $\phi$ transpose the unite mass $\mu=e^t$ 
of the characters $\phi_\pm$
\i\label{01}
\phi(x+\vep t,\vep)=e^{t\vep Z_0}\phi(x,\vep)\,e^{-t\vep Z_0}
=(\phi_-^{-1}\cdot\phi_+)(x,t,\vep). \f
We define $\tphi_+:=\phi_+\,e^{t\vep Z_0}\in\tG_+$ and obtain
\i\label{02} 
e^{t\vep Z_0}\,(\phi_-^{-1}\cdot\tphi_+)(x,0)=(\phi_-^{-1}\cdot\tphi_+)(x,t).\f
Differentiating above equation by $t$, 
we obtain equation for $\tL$
\i\label{03} 
\del{\phi_-}{t}\,\phi_-^{-1}=-(\phi_-\vep Z_0\phi_-^{-1})_-,\q
\del{\tphi_+}{t}\,\tphi_+^{-1}=(\phi_-\vep Z_0\phi_-^{-1})_+ \f
where $(\cdot)_\pm$ denote the projection onto $\tL_\pm$.
We assume that there exists the limit \cite{ck2} 
$F_t(x):=\lim_{\vep\rightarrow0}(\phi_-\th_{\vep}\phi_-^{-1})$ 
and define $\b(x)\in\tL$ such that $F_t=e^{t\beta}$. 
The element $\b$ satisfies \cite{ck2}  
\i\label{031} \phi_-\left(\del{\th_t}{t}\right)_{t=0}\hspace{-5mm}(\phi_-^{-1})= 
\phi_-\vep Z_0\phi_-^{-1}-\vep Z_0 =\beta\in \tL\f
Then, equations (\ref{03}) are 
\i\label{04}
\del{\phi_-}{t}=0,\q\del{\tphi_+}{t}\,\tphi_+^{-1}=\beta+\vep Z_0. \f
Note that these equations are equivalent to
\i\label{05} 
\left(\vep \del{}{x}-\del{}{t}\right)\phi_+\cdot\phi_+^{-1}=
\left(\vep \del{}{x}-\del{}{t}\right)\phi_-\cdot\phi_-^{-1}=\beta,  
\f
since $\th_{x'}(\phi_\pm)(x)=\phi_\pm(x+x')$.
The ``Baker function'' \cite{w2} $w(x,t,\vep):=\phi_-e^{t\vep Z_0}\in\tG$ satisfies 
same equation with $\tilde{\phi}_+$.
The first equation of (\ref{04}) means the well-known fact 
that $\phi_-$ dose not depend on unit mass. 
The second equation shows that $\phi_+$ depend on the unit mass $t$ as 
\i\label{06} 
\phi_+(x,t,\vep)=e^{t(\beta+\vep Z_0)}\phi_+(x,0,\vep)\,e^{-t\vep Z_0}. 
\f
For $\vep=0$, of course, (\ref{06}) is reduced to $\phi_+(t)=e^{t\beta}\phi_+(0)$. 

The counter term $\phi_-$ does not necessarily have the form $\phi_-\!=\!e^{\alpha}$ 
with some $\alpha\in L_-$ as the integrable systems \cite{w1}.
However, it has the scattering type formula \cite{ck2} 
\i\label{07} 
\phi_-(x,\vep)=\lim_{t\lra\infty}e^{-t(\frac{\b}{\vep}+Z_0)}e^{tZ_0}.\f
Here, we give another proof than that of Ref. \cite{ck2}. 
Owing to equation (\ref{031}),
we have
\i\label{08} &&
e^{-t(\frac{\beta}{\vep}+Z_0)}e^{tZ_0}=e^{-t(\phi_-Z_0\phi_-^{-1})}e^{tZ_0}\non\\
&&\hspace{1cm}
=(\phi_- e^{-tZ_0}\phi_-^{-1})e^{tZ_0}=\phi_-\th_{-t}(\phi_-^{-1}).
\f
For $X\in\H$ with $\Delta(X)=\sum X'\otimes X''$,  we have
\i\label{09} 
&&\phi_-\th_{-t}(\phi_-^{-1})(X)=\langle\phi_-\otimes \phi_-^{-1}, \sum
X'\otimes \th_{-t}(X'')\rangle \non\\
&&\hspace{2.5cm}=\phi_-(X)\phi_-^{-1}(1)+O(e^{-t})\lra\phi_-(X)\f 
as $t\lra\infty$. This shows the formula (\ref{07}).

We can rewrite this formula in terms of the characters.
Owing to (\ref{06}) and (\ref{07}), we have
\i\label{10} 
e^{-tZ_0}e^{t(\frac{\b}{\vep}+Z_0)}=
e^{-tZ_0}\phi_+(x,t\vep^{-1},\vep)e^{tZ_0}\phi^{-1}_+(x,0,\vep).
\f
Therefore we obtain 
\i\label{11} 
\phi_-^{-1}(x,\vep)=\lim_{t\lra\infty}\phi_+(x-
t,t\vep^{-1},\vep)\phi^{-1}_+(x,0,\vep)   
\f 
and
\i\label{12} 
\phi(x,\vep)=\lim_{t\lra\infty}\phi_+(x-
t,t\vep^{-1},\vep).
\f 
This formulae imply that 
the characters $\phi,\phi_-$ are recovered from $\phi_+$ with the
suitable limits.

\section{The differential equations of the coordinate $\vep$ on $\P^1$}

We can obtain the differential equation of $\vep$.
Differentiating equation
$[Z_0,\phi_-^{-1}]=\frac{1}{\vep}\beta\phi_-^{-1}$ by $\vep$, 
we have
\i\label{121}
[Z_0,\dot{\phi}_-^{-1}\phi_-]=-\frac{1}{\vep^2}\phi_-^{-1}\b\phi_- \f
where $\dot{\phi}_-^{-1}=\del{}{\vep}(\phi_-^{-1})$.
In general, if $\a,\a'\in L$ satisfies $\a(1)\!=\!\a'(1)\!=\!0$ and 
$[Z_0,\a]=\a'$, $\a$ is give by the integral form \cite{ck2,k1}
\i \a=\int_0^\infty \!\!\!dt \;\th_{-t}(\a'). \f
Since $\phi_-(1)=1,\; \b(1)=0$, we have
$(\dot{\phi}_-^{-1}\phi_-)(1)=(\phi_-^{-1}\b\phi_-)(1)=0$. Therefore, by
equation (\ref{121}), we obtain equation
\i \del{\,\phi_-^{-1}}{\,\vep}\,\phi_-=-\frac{1}{\vep^2}M(x,\vep)  \f
where
\i M(x,\vep)=\int_{0}^\infty\!\!\!\!
dt\;\th_{-t}(\phi_-^{-1}\beta\phi_-)\in \tL. \f
If $X\in\H$ is a 1PI with the degree $n$, $M(x,\vep)(X)$ is the polynomial of
$\vep^{-1}$ whose highest degree is $n-\!1$. 
We can also derive the equation for the Baker
function $w$
\i \del{w}{\vep}\,w^{-1}=\tilde{M}(x,t,\vep),\q 
\tilde{M}:=\phi_-(M+tZ_0)\phi_-^{-1}\in\tL.  \f
Note that the differential equations of $\vep$ are used
in the theory of the monodromy preserving deformation of the 
integrable systems.
\vspace{3mm}

We can extend above results to the case of $G_2$  
which is the group of the formal diffeomorphism \cite{ck3,ck2} of $\C$
$G_2:=\{\varphi\in{\rm Diff}(\C)\,|\,\varphi(z)=z+O(z^2)\;z\!\in\!\C \}$, 
 with the help of the map \cite{ck2} $\rho:G\lra G_2$ 
which is the anti-homomorphism
$\rho(\phi_1\phi_2)=\rho(\phi_2)\circ\rho(\phi_1)$ of the group.
For example, owing to the map $\rho$ and equation (\ref{11}), we have    
\i \psi^{-1}(\vep)=\lim_{t\lra\infty}\psi_+^{-1}(0,\vep)\circ
\a_{-t}(\psi_+)(t\vep^{-1},\vep)   \f
where $\psi(\vep)=\rho(\phi_-(0,\vep)),\;
\psi_+(t,\vep)=\rho(\phi_+)(0,t,\vep)$ and 
the $\a_t$ is the grading \cite{ck2} of $G_2$ which satisfies 
$\a_t\circ\rho=\rho\circ\th_t$.

\section{The hierarchy of the renormalization group}

We can introduce the hierarchy with the left action of 
$\exp\left(\sum_{n\ge1} t_n\vep^n Z_0\right)$ on $\phi$ 
as the integrable systems \cite{w2,w1}.
These flows are simple. The $\phi_-$ does not depend on $t_n$ again and 
$\tphi_+:=\phi_+ e^{\sum_{n\ge1} t_n\vep^n Z_0}$ satisfies the
equation 
$\del{\tphi_+}{t_n}\tphi_+=\vep^{n-1}(\b+\vep Z_0)$. 
The $\{t_n\}$ parametrize the $\vep^n$ order flows of the renormalization.

If we enlarge the group $G$ or $G_2$, the hierarchy generates the 
non-trivial flows.
For example, we choose the Lie group whose Lie algebra $\cal{L}$ is generated by 
$\{Z_{-1},Z_0,\{Z_T\}_T\}$ where $T$ is the rooted tree   
in the sense of Ref. \cite{ck3}, 
the hierarchy is the AKNS type one associated with $\cal{L}$.
In fact, we restrict $\cal{L}$ to the Lie subalgebra ${\mathfrak sl}(2,\C)$
of ${\cal L}$ generated by $\{Z_{-1},Z_0,Z_\bullet\}$ and the hierarchy 
corresponds to the usual AKNS hierarchy \cite{w1}.


\vspace*{6pt}

\begin{thebibliography}{0}
\bibitem{ck3} A. Connes and D. Kreimer, {\it Commun. Math. Phys.} {\bf
	199} 203 (1998). 
\bibitem{ck1} A. Connes and D. Kreimer, {\it JHEP} {\bf 9909}, 024 (1999). 
\bibitem{ck2} A. Connes and D. Kreimer, {\it Commun. Math. Phys.} {\bf
	216} 215 (2001). 
\bibitem{k1} D. Kastler, hep-th/0104017.
\bibitem{w2} G. Wilson, {\it C. R. Acad. Sci. Paris} {\bf 299} 587 (1984).
\bibitem{w1} G. Wilson, in {\it Integrable Systems: The Verdier memorial
	conference : actes du colloque international de Luminy}, %(Progress in Mathematics  Vol. 115)
        eds.~O. Babelon {\it et al.}  (Birkh\"auser, 1993).
\end{thebibliography}
\end{document}